# Axisymmetric Gravitational Solutions As Possible Classical Backgrounds Around Closed String Mass Distributions

Hitoshi NISHINO[1] and Subhash RAJPOOT[2]

*Department of Physics & Astronomy*
*California State University*
*Long Beach, CA 90840*

## Abstract

By studying singularities in stationary axisymmetric Kerr and Tomimatsu-Sato solutions with distortion parameter $\delta = 2, \ 3, \ \cdots$ in general relativity, we conclude that these singularities can be regarded as nothing other than closed string-like circular mass distributions. We use two different regularizations to identify $\delta$-function type singularities in the energy-momentum tensor for these solutions, realizing a regulator independent result. This result gives supporting evidence that these axisymmetric exact solutions may well be the classical solutions around closed string-like mass distributions, just like Schwarzschild solution corresponding to a point mass distribution. In other words, these axisymmetric exact solutions may well provide the classical backgrounds around closed strings.

---

[1]E-Mail: hnishino@csulb.edu

[2]E-Mail: rajpoot@csulb.edu

## 1. Introduction

Since its first discovery in general relativity in 1916 [1], it has been well understood that the Schwarzschild solution describes the classical gravitational background around a point mass. If there is a mathematically rigorous 'point mass' in reality, it is natural to expect a basic and simple classical solution in general relativity that describes the space-time around such a 'point mass'. On the other hand, developments in string theory [2] tell us that such a 'point mass' description of a particle is unrealistic due to divergent self energy, but instead, it must be replaced by a more 'smoothed' one, like a circular mass distribution in closed string theory [2].

However, if that is really the case, and such a circular mass distribution is a fundamental physical entity, it seems unnatural that no basic exact classical solution has ever been discovered as the fundamental classical background around such a circular mass distribution, as an improved version of Schwarzschild solution, or as more physically realistic mass distributions. For example, Chazy-Curzon solution [3] gives a simplest metric around a static axisymmetric mass distribution. However, its actual mass distribution is not at a finite radius, but is confined to the origin as a certain limit. What is missing is the classical background solution around a circular closed string like mass distribution with a finite radius.

The idea of studying classical solutions in general relativity around singular mass distributions is also motivated by the recent developments in brane theory known as braneworld scenario [4][5], or Randall-Sundrum scenario [6] in which the matter distributions are limited to the four-dimensional (4D) brane hypersurface embedded into 5D with $\delta$-function type singularities. Such singularities on submanifolds or boundaries play an important role in brane theories, inducing a special effect such as desirable mass hierarchies crucial to phenomenology in 4D [6], or embedding 10D supergravity into 11D [4]. As a matter of fact, some ideas relating the singularities in the Kerr solution in general relativity [7] to strings [2] have been proposed in [8], or a new concept of Bekenstein-Hawking black hole entropy interpreted microscopically in terms of elementary string excitations has been presented [9].

From these developments, it seems natural to consider the physical significance of the singularities in stationary axisymmetric exact solutions in vacuum in general relativity, in particular, their possible link with singular mass distributions that resemble closed strings. Some of these exact solutions may well be analogous to Schwarzschild solution [1] that corresponds to a point mass distribution. In our present paper, we try to show that the series of axisymmetric solutions in vacuum, starting with the Kerr solution [7] and the Tomimatsu-Sato (TS) solutions [10][11] with the distortion parameter $\delta \geq 2$ seem to describe nothing other than the classical backgrounds around circular mass distributions with coaxial radii with naked singularities for $q > 1$ [10], resembling closed strings themselves. To put it differently, we try to show that the naked singularities in TS solutions with $\delta \geq 2$ [10][11] can be interpreted as closed string-like mass distributions. As a methodology, we adopt regularization schemes similar to that in [12], *i.e.*, we use Yukawa-potential type regularizations with an exponential damping factor, that give the right coefficients for the
$\delta$-function like mass distributions. In ref. [12], this regularization was applied successfully to
the cases of Schwarzschild and Kerr solutions. In our present paper, we generalize this result
to the more general case of TS solutions with the distortion parameter $\delta = 2,\ 3,\ \cdots$ [10][11].
We use two types of regulators: The first one is equivalent to that in [12], while the second
one is its slight modification. By comparing the results of these two regulators, we will
demonstrate no regulator dependence in our results, which may be taken as the correctness
of our physical interpretation of the ring singularities as closed string-like circular mass
distributions.

## 2. A Preliminary with Circular Charge Distribution for Coulomb Potential

We first analyze the case of circular charge distribution for the case of an electric field with
a Coulomb potential, as a preliminary for the later case of TS solutions in general relativity.
Consider a circular charge distribution at radius $\rho = a$ on the equatorial $xy$-plane with
constant line charge density $\sigma$ in cylindrical coordinates $(\rho, \varphi, z)$. We need to get the
Coulomb potential $\phi(\rho, \varphi, z)$ around such a distribution, satisfying the Laplace equation
$\Delta_3 \phi = 0$. Next, given such a result for $\phi(\rho, \varphi, z)$, we want to reconstruct the original charge
distribution, which should contain $\delta(\rho - a)\,\delta(z)$. The latter scenario is highly non-trivial,
because if we blindly substitute the result for $\phi(\rho, \varphi, z)$ back into the Laplace equation,
we get zero everywhere, by definition. What is required here is the careful treatment of the
singularity at $\rho = a$ and $z = 0$. In other words, we need a good regularization to handle
the singularity.

There are several regularizations to treat such singularities. In this paper, we adopt finite
mass regularizations, based on the 3D 'massive' Laplace equation

$$\Delta_3 \phi - \kappa^2 \phi = 0 \ , \tag{2.1}$$

that replaces the original 'massless' Laplace equation.

The Coulomb potential $\phi(\rho, \varphi, z)$ around the circular charge distribution can be obtained
by the use of a massive Green's function of the Yukawa-potential type $e^{-\kappa r}/r$, where $r =
[(\rho \cos\varphi - a\cos\vartheta)^2 + (\rho\sin\varphi - a\sin\vartheta)^2 + z^2]^{1/2}$ is the distance from the charge between
the azimuthal angles $\vartheta \sim \vartheta + \delta\vartheta$ on the circle $\rho = a$ on the $xy$-plane and the point
$(\rho\cos\varphi, \rho\sin\varphi, z)$. Let us call such a small contribution $\delta\phi$ to the total potential $\phi$. The
total potential $\phi$ is obtained by superimposing $\delta\phi$ over $0 \leq \vartheta < 2\pi$ at $\rho = a$:

$$\phi(\rho, \varphi, z) = \sum \delta\phi = \int d\phi = \int_0^{2\pi} d\vartheta \, \frac{\sigma a e^{-\kappa\sqrt{(\rho\cos\varphi - a\cos\vartheta)^2 + (\rho\sin\varphi - a\sin\vartheta)^2 + z^2}}}{\sqrt{(\rho\cos\varphi - a\cos\vartheta)^2 + (\rho\sin\varphi - a\sin\vartheta)^2 + z^2}}$$

$$= \sigma a \int_0^{2\pi} d\vartheta \, \frac{e^{-\kappa\sqrt{\rho^2 + z^2 + a^2 - 2a\rho\cos\vartheta}}}{\sqrt{\rho^2 + z^2 + a^2 - 2a\rho\cos\vartheta}} \ , \tag{2.2}$$

where the $\varphi$-dependence has disappeared in the final expression, because of the periodicity of the original integrand under $\vartheta - \varphi \to \vartheta - \varphi + 2\pi$, also consistent with the axial symmetry. Due to its complicated nature, the final result in a closed form is more involved than elliptic integrals, and thus the $\vartheta$-integral here is not analytically performed.

Given such a Coulomb potential $\phi(\rho, z)$ in (2.2), our next question is whether we can 'reconstruct' the original charge distribution singular at $\rho = a$, $z = 0$. As has been mentioned, putting simply $\kappa = 0$ from the outset will lead to a vanishing result by definition. One method to avoid this is to compute $\int d^3\vec{x}\, \Delta_3 \phi$ by equating it with $\int d^3\vec{x}\, \kappa^2 \phi$, under the massive Laplace equation $\Delta_3 \phi - \kappa^2 \phi = 0$:

$$I(\kappa) \equiv \int d^3\vec{x}\, \kappa^2 \phi = \sigma a \int d^3\vec{x} \int_0^{2\pi} d\vartheta\, \kappa^2 \frac{e^{-\kappa\sqrt{\rho^2+z^2+a^2-2a\rho\cos\vartheta}}}{\sqrt{\rho^2+z^2+a^2-2a\rho\cos\vartheta}}$$

$$= \sigma a \kappa^2 \int_0^\infty d\rho \int_{-\infty}^\infty dz \int_0^{2\pi} d\varphi \int_0^{2\pi} d\vartheta\, \rho \frac{e^{-\kappa\sqrt{\rho^2+z^2+a^2-2a\rho\cos\vartheta}}}{\sqrt{\rho^2+z^2+a^2-2a\rho\cos\vartheta}} \quad . \quad (2.3)$$

The question now is whether or not the integral $I(\kappa)$ in (2.3) gives a non-vanishing finite result that corresponds to the right total charge $q \equiv 2\pi a \sigma$, after the limit $\kappa \to 0^+$. To this end, we first change the coordinate variables from $(\rho, \varphi, z)$ to $(r, \theta, \varphi)$, such that its Jacobian equals the familiar value $r$:

$$I(\kappa) = 2\pi\sigma a \kappa^2 \int_0^{2\pi} d\vartheta \int_0^\infty dr \int_0^\pi d\theta\, r \frac{e^{-\kappa\sqrt{r^2+a^2\sin\vartheta}}}{\sqrt{r^2+a^2\sin^2\vartheta}} (r\cos\theta + a\cos\vartheta)$$

$$= 2\pi\sigma a \kappa^2 \int_0^{2\pi} d\vartheta \int_0^\infty dr\, \frac{r\, e^{-\kappa\sqrt{r^2+a^2\sin\vartheta}}}{\sqrt{r^2+a^2\sin^2\vartheta}} (2r + 2\pi a\cos\vartheta) \quad . \quad (2.4)$$

Here we have performed the $\theta$-integral. We have developed in [12] the following lemma about the $r$-integration such as in (2.4) for $m \leq +1$:

$$\kappa^2 \int_0^\infty d\eta\, (\eta+1)^m\, e^{-\kappa\eta} \to \begin{cases} 0 & (m = 0,\, -1,\, -2,\, \cdots) \\ 1 & (m = +1) \end{cases}, \quad (\text{as } \kappa \to 0^+) \quad . \quad (2.5)$$

To use this lemma in our case, we use a new variable $\eta \equiv \sqrt{r^2 + b^2} - b$ $(b \equiv a\sin\vartheta)$, and expand the integrand around $\eta = -1$. Thus the first term in (2.4) is

$$2\pi\sigma a \kappa^2 \int_0^{2\pi} d\vartheta \int_0^\infty dr\, \frac{2r^2 e^{-\kappa\sqrt{r^2+b^2}}}{\sqrt{r^2+b^2}} = 4\pi\sigma a \kappa^2 \int_0^{2\pi} d\vartheta \int_0^\infty d\eta\, (\eta+b) \frac{\sqrt{\eta^2+2b\eta}}{\eta+b} e^{-\kappa(\eta+b)}$$

$$= 4\pi\sigma a \kappa^2 \int_0^{2\pi} d\vartheta\, e^{-\kappa b} \int_0^\infty d\eta\, (\eta+1) \left[1 + \frac{b-1}{\eta+1} + \mathcal{O}((\eta+1)^{-2})\right] e^{-\kappa\eta}$$

$$= 4\pi\sigma a \int_0^{2\pi} d\vartheta\, e^{-\kappa a\sin\vartheta} + \mathcal{O}(\kappa) = 4\pi\sigma a \int_0^{2\pi} d\vartheta\, (1 - \kappa a\sin\vartheta) + \mathcal{O}(\kappa)$$

$$\to 8\pi^2 \sigma a \quad (\text{as } \kappa \to 0^+) \quad . \quad (2.6)$$

Similarly, the second term in (2.4) can be shown to be zero after $\kappa \to 0^+$, and therefore we get the desirable result

$$I(\kappa) \quad \to \quad 8\pi^2 \sigma a = 4\pi(2\pi a \sigma) = 4\pi q \ , \tag{2.7}$$

for the total charge $q \equiv 2\pi a \sigma$ on the circle.

Since we have obtained a non-vanishing finite result for $I(\kappa)$, it is natural to identify the integrand with a combination of $\delta$-function such as $\delta(\rho - a)\delta(z)$. However, there are a few caveat for such an identification. For example, we have to make sure that the integrand before the $\int d^3\vec{x}$-integration vanishes almost everywhere except for $(\rho, z) = (a, 0)$. We can itemize such criterions for the identification with the $\delta$-functions, as follows, by letting $f(\rho, z; \kappa)$ be the integrand in $I(\kappa) \equiv \int d^3\vec{x} f(\rho, z; \kappa)$:

$$\text{(i)} \quad \lim_{\kappa \to 0^+} f(\rho, z; \kappa)\Big|_{\rho \neq a, \ z \neq 0} = 0 \ , \tag{2.8}$$

$$\text{(ii)} \quad \lim_{\kappa \to 0^+} \lim_{\rho \to a} f(\rho, z; \kappa)\Big|_{z \neq 0} = 0 \ , \tag{2.9}$$

$$\text{(iii)} \quad \lim_{\kappa \to 0^+} \lim_{z \to 0} f(\rho, z; \kappa)\Big|_{\rho \neq 0} = 0 \ , \tag{2.10}$$

$$\text{(iv)} \quad \lim_{\kappa \to 0^+} \lim_{\rho \to a, \ z \to 0} f(\rho, z; \kappa) = \infty \ , \tag{2.11}$$

$$\text{(v)} \quad \lim_{\kappa \to 0^+} \int d^3\vec{x} f(\rho, z; \kappa) = 8\pi^2 \sigma a \ . \tag{2.12}$$

Eq. (2.8) is to guarantee that the integrand is almost everywhere zero, while (2.9), (2.10) and (2.11) imply that the integrand has a singularity only at points satisfying both $\rho = a$ and $z = 0$, i.e., the circle with the radius $a$ from the origin on the equatorial $xy$-plane. Eq. (2.12) has been already satisfied by (2.6). Now, (2.8) is manifestly satisfied, because there is no other singularities in $f(\rho, z; \kappa)$ other than the points on $(\rho, z) = (a, 0)$. For the same reason, eqs. (2.9) and (2.10) are also satisfied. The remaining one (2.11) can be also confirmed as

$$\lim_{\rho \to a, \ z \to 0} f(\rho, z; \kappa) = 4\kappa^2 \int_0^{\pi/4} dx \, \frac{e^{-\kappa a \sin x}}{2a \sin x} = \infty \ , \tag{2.13}$$

for $x \equiv \vartheta/2$. This is due to the logarithmic divergence at $x \to 0$.

Since the criterion (i) through (v) are satisfied, we can identify

$$\lim_{\kappa \to 0^+} \kappa^2 \rho \int_0^{2\pi} d\vartheta \, \frac{e^{-\kappa\sqrt{\rho^2 + z^2 + a^2 - 2a\rho\cos\vartheta}}}{\sqrt{\rho^2 + z^2 + a^2 - 2a\rho\cos\vartheta}} = 4\pi \, \delta(\rho - a)\, \delta(z) \ . \tag{2.14}$$

## 3. Ring Singularities and Circular Mass Distributions for TS-Solutions

Once we have understood the basic case of circular charge distribution for Coulomb potential, it is much easier to handle similar singularities in TS solutions [10][11]. As has been mentioned, we will adopt the previous method in [12], which is also consistent with the criterions above.

The Kerr metric [7] is classified as a generalization of the Schwarzschild metric [1] due to rotation, while Weyl metric [13] is another generalization due to deformation. The generalizations due to both rotation and deformation yield TS metrics [10][11] with the distortion parameter $\delta = 2, 3, \cdots$. In particular, the case $\delta = 1$ of the TS metric [10] is equivalent to the Kerr metric [7], while in the cases of $\delta \geq 2$ there are naked ring singularities lying outside non-singular event horizons. These 'naked' ring singularities outside event horizons serve as actual circular mass distributions in our work.

We start with the metric of axisymmetric TS solutions with the general distortion parameter $\delta = 1, 2, \cdots$ [10]:

$$ds^2 = -e^{2\nu}(dt)^2 + m^2 e^{2\psi}(d\varphi - \Omega dt)^2 + m^2 e^{2\mu_2}(dx)^2 + m^2 e^{2\mu_3}(dy)^2 , \tag{3.1}$$

$$e^{2\nu} \equiv \frac{p^2(x^2-1)B}{D} , \quad e^{2\psi} \equiv \frac{(1-y^2)D}{\delta^2 B} , \quad \Omega \equiv -\frac{2\delta^2 qC}{mD} ,$$
$$e^{2\mu_2} \equiv \frac{B}{\delta^2 p^{2\delta-2}(x^2-1)(x^2-y^2)^{\delta^2-1}} , \quad e^{2\mu_3} \equiv \frac{B}{\delta^2 p^{2\delta-2}(1-y^2)(x^2-y^2)^{\delta^2-1}} , \tag{3.2}$$

where $A$, $B$, $C$ and $D$ are respectively the polynomials of $x$ and/or $y$ with the respective degrees $2\delta^2$, $2\delta^2$, $2\delta^2 - 1$ and $2\delta^2 + 2$.

The ring singularities exist as the zeros of the algebraic equation $B = 0$ [10][11], lying on the $y = 0$ equatorial plane at finite coaxial radii $x = x_1, x_2, \cdots, x_\delta$ which are the zeros of $B|_{y=0} = 0$. Our task is to show that the corresponding energy-momentum tensor $T^{00}$ have $\delta$-function type ring singularities at radii $x = x_1, \cdots, x_\delta$, on the equatorial plane at $y = 0$.

To this end, we look into the corresponding Einstein tensor $G^{00} \equiv R^{00} - (1/2)g^{00}R$, with an appropriate regularization like that in [12]. The 'regularized' total energy or mass for such an exact solution is given in a closed form by [14]

$$\begin{aligned} P^0 &= \int \sqrt{-g}\, T^{00}\, dS_0 = \int \sqrt{-g}\, e^{-\nu} e_{(0)}{}^0 \frac{1}{8\pi}\left(R^{(0)(0)} + \tfrac{1}{2}R\right) dS_0 \\ &= \frac{m}{4}\int_{-1}^{1} dy \int_{x_0}^{\infty} dx \Big[ -e^{\psi-\nu}\{e^{-\widetilde{\mu}_2}(e^{\widetilde{\mu}_3})_x\}_x - e^{\psi-\nu}\{e^{-\widetilde{\mu}_3}(e^{\widetilde{\mu}_2})_y\}_y \\ &\qquad - e^{-\nu}\{e^{-\widetilde{\mu}_3-\widetilde{\mu}_2}(e^{\psi})_x\}_x - e^{-\nu}\{e^{-\widetilde{\mu}_2-\widetilde{\mu}_3}(e^{\psi})_y\}_y \\ &\qquad - \tfrac{1}{4}e^{3\psi-\widetilde{\mu}_2+\widetilde{\mu}_3-3\nu}\Omega_x^2 - \tfrac{1}{4}e^{3\psi+\widetilde{\mu}_2-\widetilde{\mu}_3-3\nu}\Omega_y^2 \Big] . \end{aligned} \tag{3.3}$$

The $x_0$ is the lower limit of the $x$-integration [10]. All subscripts of coordinates denote differentiations with respect to the coordinates, e.g., $(e^\psi)_y \equiv \partial_y e^\psi \equiv \partial e^\psi / \partial y$ [14]. Following ref. [12], we have regularized the functions $\mu_2$ and $\mu_3$ by a Yukawa-potential type damping factor, denoted by $\widetilde{\mu}_2$ and $\widetilde{\mu}_3$:

$$e^{\widetilde{\mu}_2} \equiv e^{\mu_2} S , \quad e^{\widetilde{\mu}_3} \equiv e^{\mu_3} S , \quad S \equiv 1 + b\,\alpha\, e^{-\alpha\xi} , \tag{3.4}$$

where $b$ is a non-zero real constant independent of $\alpha > 0$, to be fixed shortly. Since we are interested in the case $a > m$ (equivalently $q > 1$) to see naked singularities in TS solutions, we use also $x = +i\hat{x}$, $p = -i\hat{p}$ appropriately [10][11][15]. The $\xi$ in (3.4) is defined by $\xi \equiv \hat{x} - \hat{x}_0 \equiv -i(x - x_0)$ with the range $0 \leq \xi < \infty$. We use in this paper the exponent $-\alpha\xi$ in the regulator mimicking the previous Coulomb case (2.2) or as in [12]. However, in the next section, we will use a slightly modified regulator, in order to see possible regulator dependence. If we take the limit $\alpha \to 0^+$ before the $x$ and $y$-integrations, we simply get zero, satisfying the basic criterion like (2.8) in the Coulomb case.

Since the regulator $S$ in (3.4) is inserted only into $e^{\mu_2}$ and $e^{\mu_3}$, the effect of the regulator in the integrand will arise only when the derivatives $\partial_x$ and/or $\partial_y$ hit the regulator $S$ at least once:

$$\begin{aligned}
P^0 &= + \frac{m}{4} \int_{-1}^{1} dy \int_{\xi_0}^{\infty} d\xi \left[ + e^{\psi-\nu} \{e^{-\tilde{\mu}_2}(e^{\tilde{\mu}_3})_\xi\}_\xi - e^{\psi-\nu} \{e^{-\tilde{\mu}_3}(e^{\tilde{\mu}_2})_y\}_y \right] \\
&= + \frac{m}{4} \int_{-1}^{1} dy \int_{0}^{\infty} d\xi \left[ -(e^{\psi-\nu})_\xi e^{-\mu_2} S^{-1}(e^{\mu_3} S)_\xi + (e^{\psi-\nu})_y e^{-\mu_3} S^{-1}(e^{\mu_2} S)_y \right] \\
&\quad + \frac{m}{4} \int_{-1}^{1} dy \left[ e^{\psi-\mu_2-\nu} S^{-1}(e^{\mu_3} S)_\xi \right]_{\xi=0}^{\xi=\infty} - \frac{m}{4} \int_{0}^{\infty} d\xi \left[ e^{\psi-\mu_3-\nu} S^{-1}(e^{\mu_2} S)_y \right]_{y=-1}^{y=1} . \quad (3.5)
\end{aligned}$$

In the first term we have a sign flip caused by the imaginary unit in $\xi \equiv \hat{x} - \hat{x}_0 \equiv -i(x - x_0)$. We have also performed partial integrations under the $x$ and $y$-integrations. The last two terms on the boundaries have no contributions when $\alpha \to 0^+$, since they are at $\mathcal{O}(\alpha^2)$. The second term in (3.5) is seen to be zero because at least one derivative $\partial_y$ should hit $S$ which in turn is independent of $y$, yielding a zero contribution. This is because unless there is a $y$-derivative on $S$, there will be a cancellation between $S^{-1}$ and $S$, which is exactly the same as the non-regularized case.

The only term remaining in (3.5) is

$$P^0 = -\frac{m}{4} \int_{-1}^{1} dy \int_{0}^{\infty} d\xi \left[ \frac{1}{\delta p} \sqrt{\frac{1-y^2}{x^2-1}} \frac{D}{B} \right]_\xi \sqrt{\frac{x^2-1}{1-y^2}} S^{-1} S_\xi , \quad (3.6)$$

which can be shown to be nonzero and finite. To this end, we need a special lemma for the factor $e^{\psi-\nu}$ involving $B$ and $D$ for a general $\delta$:

$$B = p^{2\delta} x^{2\delta^2} + \mathcal{O}(x^{2\delta^2-1}) , \quad D = p^{2\delta+2} x^{2\delta^2+2} + \mathcal{O}(x^{2\delta^2+1}) , \quad (3.7)$$

justified by the solution for a general $\delta$ for the polynomial $A$ [10]:

$$A \equiv |\alpha|^2 - |\beta|^2 = 2 \sum_{r=1}^{\delta} (-1)^{r+1} d_{r+1} f_r c_{\delta,r} F_r = \frac{\mathcal{D}}{\Delta} , \quad (3.8a)$$

$$F_r \equiv \frac{\mathcal{D}_r}{\Delta_r} , \quad \mathcal{D} \equiv \det\left(\frac{f_{r+r'-1}}{r+r'-1}\right) , \quad \Delta \equiv \det\left(\frac{1}{r+r'-1}\right) , \quad (3.8b)$$

$$c_{\delta,r} \equiv 2^{2r-1} \frac{\delta(\delta+r-1)!}{(\delta-r)!(2r)!} , \quad d_r \equiv \frac{(2r-3)!!}{(2r-2)!!} , \quad f_r \equiv p^2(x^2-1)^r + q^2(y^2-1)^r , \quad (3.9)$$

where $\mathcal{D}_r$ (or $\Delta_r$) is the co-factor of the determinant $\mathcal{D}$ (or $\Delta$) for the $(1, r)$ component.

We can prove that only the term with $\mathcal{D}_1$, among $\delta$ terms in the sum in (3.8), has the highest power in $x$: $\mathcal{D}_1 \approx p^{2\delta-2} x^{2\delta^2 - 2} + \mathcal{O}(x^{2\delta^2 - 3})$. Note that $\mathcal{D}_r$ is proportional to

$$\mathcal{D}_r = \frac{1}{(\delta - 1)!} \epsilon^{i_1 \cdots \widehat{i_r} \cdots i_\delta} \mathcal{D}_{1\,i_1} \mathcal{D}_{2\,i_2} \cdots \widehat{\mathcal{D}_{r\,i_r}} \cdots \mathcal{D}_{\delta\,i_\delta}$$
$$\approx \epsilon^{i_1 \cdots \widehat{i_r} \cdots i_\delta} \mathcal{O}\left((x^2)^{(1+2+\cdots+\widehat{r}+\cdots+\delta)+(i_1+i_2+\cdots+\widehat{i_r}+\cdots+i_\delta)-(\delta-1)}\right) , \tag{3.10}$$

where $\mathcal{D}_{ij}$ is the $ij$-component of the determinant matrix in (3.8b), e.g., $\mathcal{D}_{2\,i_2} \approx \mathcal{O}\left(p^2(x^2 - 1)^{2+i_2-1} + q^2(y^2 - 1)^{2+i_2-1}\right) \approx \mathcal{O}(x^{2+i_2-1})$, etc., while a 'hatted' factor or term is to be skipped in the product or sum. Now it is clear that the highest power of $x$ arises only in the case of $r = 1$ with $i_1 = 1$, so that

$$\mathcal{D}_1 \approx (p^2)^{\delta-1} (x^2)^{2(\delta-1)(\delta-2)/2 - \delta + 1} \approx p^{2\delta-2} x^{2\delta^2 - 2} . \tag{3.11}$$

Therefore, only the term with $F_1$ in (3.8a) has the highest power:

$$A \approx p^{2\delta} x^{2\delta^2} + \mathcal{O}(x^{2\delta^2 - 1}) . \tag{3.12}$$

Recalling other relevant relationships [10][11], such as

$$A \equiv |\alpha|^2 - |\beta|^2 = u^2 + v^2 - m^2 - n^2 , \qquad B \equiv (u+m)^2 + (v+n)^2 ,$$
$$\xi \equiv \frac{\alpha}{\beta} , \qquad \alpha \equiv u + iv , \qquad \beta \equiv m + in \qquad (u, v, m, n \in \mathbb{R}) ,$$
$$G \equiv |\beta|^2 = \sum_{r=1}^{\delta} c_{\delta, r} F_r , \qquad DA = p^2(x^2 - 1) B^2 - 4\delta^2 q^2 (1 - y^2) C^2 , \tag{3.13}$$

we see that

$$G \approx c_{\delta,1} F_1 \approx \delta^2 p^{2\delta-2} x^{2\delta^2 - 2} + \mathcal{O}(x^{2\delta^2 - 3}) \approx |\beta|^2 \quad \Longrightarrow \quad \beta \approx \delta p^{\delta-1} x^{\delta^2 - 1} \approx m ,$$
$$A \approx u^2 - m^2 \approx p^{2\delta} x^{2\delta^2} \quad \Longrightarrow \quad u \approx p^\delta x^{\delta^2} , \tag{3.14}$$
$$B \approx (u + m)^2 \approx p^{2\delta} x^{2\delta^2} , \qquad D \approx \frac{p^2 x^2 B^2}{A} \approx p^{2\delta+2} x^{2\delta^2 + 2} , \tag{3.15}$$

yielding (3.7). The symbol $\approx$ denotes the omission of lower power of $x$.

We now use these leading terms for the evaluation of (3.6):

$$P^0 = \frac{bm\widehat{p}\alpha^2}{4\delta} \int_{-1}^{1} dy \int_0^\infty d\xi \sqrt{(\xi + \xi_0)^2 + 1} \, e^{-\alpha \xi} [1 + \mathcal{O}(\alpha)]$$
$$= \frac{bm\widehat{p}\alpha^2}{4\Delta} \int_{-1}^{1} dy \int_0^\infty d\xi \left[ (\xi + 1) + \mathcal{O}\left((\xi + 1)^0\right) \right] e^{-\alpha \xi} + \mathcal{O}(\alpha) \; \to \; \frac{bm\widehat{p}}{2\delta} , \tag{3.16}$$

as $\alpha \to 0^+$, where we have relied on the lemma (2.5). This gives the total mass in (3.5)

$$\lim_{\alpha \to 0^+} P^0 = m , \tag{3.17}$$

agreeing with the asymptotic mass [16] after the identification

$$b = \frac{2\delta}{\widehat{p}} \quad . \tag{3.18}$$

Our remaining task is to confirm the criterions for identification of the integrand with a desirable combination of $\delta$-functions, like (2.8) - (2.12) in the Coulomb case. Our present ones for the integrand $g(x, y; \alpha)$ under $\int dy \int dx$ in (3.3) can be dictated as

$$\text{(i)} \quad \lim_{\alpha \to 0^+} g(x, y; \alpha)\Big|_{x \neq x_i, \, y \neq 0} = 0 \, , \tag{3.19}$$

$$\text{(ii)} \quad \lim_{\alpha \to 0^+} \lim_{x \to x_i} g(x, y; \alpha)\Big|_{y \neq 0} = 0 \, , \tag{3.20}$$

$$\text{(iii)} \quad \lim_{\alpha \to 0^+} \lim_{y \to 0} g(x, y; \alpha)\Big|_{x \neq x_i} = 0 \, , \tag{3.21}$$

$$\text{(iv)} \quad \lim_{\alpha \to 0^+} \lim_{x \to x_i, \, y \to 0} g(x, y; \alpha) = \infty \, , \tag{3.22}$$

$$\text{(v)} \quad \lim_{\alpha \to 0^+} \int d^3\vec{x} \, g(x, y; \alpha) = (\text{finite and non-zero}) \, . \tag{3.23}$$

As before, (3.19), (3.20) and (3.21) are easily seen to be satisfied, while (3.23) has been already confirmed in (3.16), leaving (3.22). The only non-trivial singularity at $(x, y) \to (x_i, 0)$ $(i = 1, 2, \cdots, \delta)$ in (3.22) must arise from the first term in (3.5) out of its four terms, because only this term had non-vanishing contribution at $\alpha \to 0^+$:

$$-\frac{m}{4}(e^{\psi-\nu})_\xi \, e^{-\mu_2} \, S^{-1}(e^{\mu_3}S)_\xi = \frac{m}{4\delta\widehat{p}} \left(\frac{D}{\sqrt{x^2-1}\,B}\right)_\xi \frac{\sqrt{x^2-1}\,b\alpha^2\,e^{-\alpha\xi}}{1+b\alpha e^{-\alpha\xi}}\bigg|_{y=0} , \tag{3.24}$$

after the limit $y \to 0$ is taken. Now the question is about the behaviour of the combination $D/B$ at $y = 0$. To answer this question, we have to postulate the general polynomial structures of $A$, $B$ and $C$ at $y = 0$ as

$$A|_{y=0} = (x - x_1)(x - x_2) \cdots (x - x_\delta) \, A_{2\delta^2 - \delta}(x) \, ,$$
$$B|_{y=0} = (x - x_1)^2(x - x_2)^2 \cdots (x - x_\delta)^2 \, B_{2\delta^2 - 2\delta}(x) \, ,$$
$$C|_{y=0} = (x - x_1)(x - x_2) \cdots (x - x_\delta) \, C_{2\delta^2 - \delta - 1}(x) \, . \tag{3.25}$$

The $A_{2\delta^2-\delta}(x)$, $B_{2\delta^2-2\delta}(x)$ and $C_{2\delta^2-\delta-1}(x)$ are respectively polynomials of $x$ whose degrees are denoted by their subscripts, and they are supposed to have no zeros at $x = x_i$ $(i = 1, 2, \cdots, \delta)$. These postulates are based on the following facts [10]. First, the space-time singularities causing the Riemann tensor to blow up, occurs when two conditions $u + m = 0$ and $y = 0$ are met [10]. In particular, there are only $n$ zeros at $x = x_i$ $(i = 1, 2, \cdots, \delta)$ for the equation $u + m = 0$. Second, in terms of $\alpha = u + iv$, $\beta = m + in$ as in [10], $v = n = 0$ at $y = 0$. Third, the polynomial $(u + m)$ at $y = 0$ is supposed to have simple zeros at $x = x_i$, and therefore $B \approx (u+m)^2$ has double zeros at $x = x_i$. Fourth, $A \equiv (u+m)(u-m) + (v+n)(v-n)$ has simple zeros at $x = x_i$, due to the structure $(u+m)(u-m)$. Fifth, we assume the form of $C$ above, based on the explicit cases of $\delta = 1$ and $\delta = 2$ [10].

Once we specify the structures (3.25), it follows from (3.13) that

$$\begin{aligned} D|_{y=0} = & +p^2(x^2-1)(x-x_1)^3\cdots(x-x_\delta)^3 \frac{(B_{2\delta^2-2\delta})^2}{A_{2\delta^2-\delta}} \\ & -4\delta^2 q^2(x-x_1)\cdots(x-x_\delta)\frac{(C_{2\delta^2-\delta-1})^2}{A_{2\delta^2-\delta}} \quad . \end{aligned} \qquad (3.26)$$

Obviously there are simple poles in the combination $D/B|_{y=0}$ at $x = x_i$, because the second term here will generate simple poles after being divided by $B$ in (3.25). Therefore at least one of such poles remains after the differentiation $\partial_\xi = \partial_x$ in $(D/B)_\xi$ in (3.24). This establishes the singularity (3.22), and therefore we have the satisfaction of all the criterions for the identification of the integrand $g(x, y; \alpha)$ with the combination of the $\delta$-functions:

$$T^{00} = \frac{m}{2\pi\delta\sqrt{-g}} \sum_{i=1}^{\delta} \delta(\widehat{x} - \widehat{x}_i)\,\delta(y) \quad . \qquad (3.27)$$

The previous case for Kerr solution in [12] can be now re-obtained as the special case of $\delta = 1$. Recall that the coordinates $(\widehat{x}, y)$ are related to the familiar Weyl's canonical coordinates $(\rho, z)$ [10][15] as

$$\rho = \frac{m\widehat{p}}{\delta}\sqrt{(\widehat{x}^2+1)(1-y^2)} \quad , \qquad z = \frac{m\widehat{p}}{\delta}xy \quad , \qquad (3.28)$$

so that (3.27) implies that we have the ring singularities at the radii $\rho = \rho_i \equiv m\widehat{p}\sqrt{\widehat{x}_i^2+1}/\delta$ $(i = 1, \cdots, \delta)$ corresponding to $\widehat{x} = \widehat{x}_i$ on the equatorial plane at $z = 0$ or equivalently $y = 0$, as desired.

To turn the table around, our result can be re-interpreted in the following way: Suppose we are provided with the closed string-like multiple circular mass distributions (3.27). In total, there are $\delta$ closed strings with different radii at $\rho = \rho_i$. We can then obtain the exact classical gravitational background solution around such mass distributions, which are nothing other than the series of TS solutions with the distortion parameter $\delta$ [10][11]. In other words, our result has the significance that the classical gravitational background around multi closed strings can be given by the series of TS solutions [10][11].

## 4. Regularization Independence

We have chosen so far the particular regulator (3.4), mimicking the case of Coulomb potential (2.3). However, this does not have to be the only regulator for gravitational singularities. As a matter of fact, the exponent of a regulator does not have to be a linear function like $-\alpha\xi$, but it may be a quadratic function, such as $-\beta\xi^2$, if its purpose is just to get the finite $\xi$-integral at $\xi \to \infty$. Namely, we now consider replacing $S$ in (3.4) by

$$\widehat{S} \equiv 1 + c\beta e^{-\beta^2 \xi^2} \quad , \tag{4.1}$$

with a nonzero constant $c$, and we take the limit $0 < \beta \to 0^+$ at the end of computations. The particular power of $\beta$ in the exponent or that in the outside of the exponential function have been chosen, such that the final result is neither trivially zero nor divergent. In what follows, we briefly study the potential difference between this and the previous result (3.16), and show that there is eventually no regulator dependence.

Among the four terms in (3.5), the last two terms on the boundaries do not contribute, while the second term is also zero for reasons similar to the previous regulator. The remaining first term in (3.5) is now

$$\frac{m}{4}\int_{-1}^{1} dy \int_0^\infty d\xi \left[\frac{\widehat{p}\sqrt{1-y^2}}{\delta} + \mathcal{O}(\xi^{-1})\right] \sqrt{\frac{\widehat{x}^2+1}{1-y^2}}\, \frac{-2c\beta^3 \xi e^{-\beta^2\xi^2}}{1+c\beta e^{-\beta^2\xi^2}}\, [1 + \mathcal{O}(\beta)]$$

$$= \frac{m\widehat{p}c\beta^3}{2\delta}\int_0^\infty d\xi \left[ a_2(\xi+1)^2 + a_1(\xi+1) + a_0 + \frac{a_{-1}}{\xi+1} + \frac{a_{-2}}{(\xi+1)^2} + \mathcal{O}\bigl((\xi+1)^{-3}\bigr)\right] e^{-\beta^2\xi^2}$$

$$= \frac{m\widehat{p}c}{2\delta}\left(a_2 I_2 + a_1 I_1 + a_0 I_0 + a_{-1}I_{-1} + a_{-2}I_{-2} + \cdots\right) \quad . \tag{4.2}$$

Here we have expanded the integrand other then the factor $e^{-\beta\xi^2}$ in terms of $\xi+1$ instead of $\xi$ itself in order to avoid the singularity at $\xi = 0$, with the coefficients $a_2$, $a_1$, $\cdots$. The $I_n$'s are defined by

$$I_n \equiv \int_0^\infty d\zeta\, \beta^{2-n}(\zeta+\beta)^n e^{-\zeta^2} \quad , \tag{4.3}$$

with $\zeta \equiv \beta\xi$. Among these $I_n$'s, we can show that it is only $I_2$ whose limit is non-zero when $\beta \to 0^+$, similarly to the lemma (2.5):

$$I_n \to \begin{cases} 0 & (n=1,\,0,\,-1,\,-2,\,\cdots) \\ \frac{\sqrt{\pi}}{4} & (n=+2) \end{cases} \quad , \qquad (\text{as } \beta \to 0^+) \quad . \tag{4.4}$$

In fact, $I_1 \to 0$, $I_0 \to 0$ are easily proven, while we need the inequality $0 < I_{-m} \le I_{-(m-1)}$ for $(m=2,\,3,\,\cdots)$ shown as

$$0 < I_{-m} = \beta^{2+m}\int_0^\infty d\zeta\, \frac{e^{-\zeta^2}}{(\zeta+\beta)^m} = \beta^{2+(m-1)}\int_0^\infty d\zeta\, \frac{e^{-\zeta^2}}{(\zeta+\beta)^{m-1}}\left(1 - \frac{\zeta}{\zeta+\beta}\right)$$

$$\le \beta^{2+(m-1)}\int_0^\infty d\zeta\, \frac{e^{-\zeta^2}}{(\zeta+\beta)^{m-1}} = I_{-(m-1)} \quad . \tag{4.5}$$

Therefore, all the $I_{-m}$'s are bounded from above by $I_{-1}$ which in turn goes to zero:

$$0 < I_{-1} \equiv \int_0^\infty d\zeta\, \frac{\beta^3}{\zeta+\beta}\, e^{-\zeta^2} = \int_0^1 d\zeta\, \frac{\beta^3}{\zeta+\beta}\, e^{-\zeta^2} + \int_1^\infty d\zeta\, \frac{\beta^3}{\zeta+\beta}\, e^{-\zeta^2}$$

$$\le \int_0^1 d\zeta\, \frac{\beta^3}{\zeta+\beta} + \int_1^\infty d\zeta\, \frac{\beta^3}{\zeta}\, e^{-\zeta^2} = \beta^3 \ln(\beta+1) - \beta^3 \ln\beta + \beta^3 \mathcal{O}(\beta^0) \to 0 \quad , \tag{4.6}$$

as $\beta \to 0^+$. Now $0 < I_{-m} \leq I_{-(m-1)} \leq \cdots \leq I_{-1} \to 0$ implies that $\lim_{\beta \to 0^+} I_{-m} = 0$ ($m = 1, 2, \cdots$). Using this in (4.2) yields the desirable result

$$\lim_{\beta \to 0^+} P^0 = m , \qquad (4.7)$$

if the constant $c$ is normalized as $c = 8\delta/(\sqrt{\pi}\hat{p})$, since $a_2 = 1$ from (4.2).

As for the criterions (3.19) - (3.23), the most important pole structure in (3.24) out of $(\sqrt{x^2 - 1}D/B)_\xi$ remains the same, even though the last factor is now replaced by $\hat{S}^{-1}\hat{S}_\xi$. Therefore, all criterions are satisfied as in the previous regularization, leading us to the same identification with the $\delta$-functions as in (3.27).

This independence of regularizations is the reflection of the fact that the contribution to the total integral comes solely from the leading terms in the energy-momentum tensor $T^{00}$ as the integrand. This also strongly indicates the 'physical' correctness of our interpretation of the singularity as the closed string-like circular mass distributions.

## 5. Concluding Remarks

In this paper, we have shown by explicit computation, with the appropriate regulator (3.4) that there exist ring singularities in the energy-momentum tensor for general TS solutions at the coaxial radii $\rho = \rho_i$ ($i = 1, \cdots, \delta$) on the equatorial plane at $z = 0$, that can be identified with the $\delta$-function like singularity (3.27). As the guiding principle, we followed the case of circular charge distribution for Coulomb potential with a massive Yukawa-type damping factor, which also provides the criterion to be satisfied for the identification of the integrand with a combination of $\delta$-functions. Thanks to the generalization in ref. [11] of the TS solutions in [10] to $1 \leq \delta < \infty$, we have been able to generalize our result to an arbitrary number of the distortion parameter $\delta$ which corresponds to the number of ring singularities. In particular, there are naked singularities outside the event horizons in the cases $\delta \geq 2$ which may well be actual closed string-like mass distributions.

It has been well-known that the asymptotic mass for a given exact axisymmetric gravitational solution can be obtained by the asymptotic behaviour of such a solution at $r \approx \infty$ [16]. However, such a method is not explicit enough for us to identify the singularities in the solution with the closed string type mass distribution. One of the reasons is that the asymptotic behaviors are associated with global boundary effects, which are not detailed enough to describe 'microscopic' mass distributions. In other words, our methodology provides more direct and explicit links between exact solutions and mass distributions. We stress that the mass distribution in energy-momentum tensor is a 'physical entity', and it is not just an 'effective mass' observed only at infinity which is indirect or implicit.

We have also used in this paper two distinct regularizations (3.4) and (4.1), and have demonstrated that our result $P^0 = m$ is independent of a particular choice of regulators.

This is because the contributions to the total integral came from the 'leading' terms in the expansions, such as the highest power in (3.16), or the structure of zeros in (3.25) and (3.26). The analogous result in the Coulomb case also provides a justification of such a regularization based on a 'physical' consideration. In other words, our method is not a reflection of a mathematical artifact, but is a 'physical entity' with the total mass $m$. We also mention that, as is usually the case in physics, once physical interpretation is correct, a regularized finite result does not depend on the choice of regularizations.

If superstring theory [2] really describes the ultimate microscopic scales around the Planck mass, it implies the existence of singular mass distributions as 'physical objects'. These mass distributions are required to be rigorously singular, *i.e.*, there must be no smearing effect on such singularities. From this viewpoint, it is natural that the gravitational background around such singular mass distributions exist in a relatively simple and fundamental form, such as generalized TS solutions for $\delta \geq 2$ [10][11]. Our result in this paper has provided supporting evidence for such philosophy.

In the Randall-Sundrum scenario [6], the 4D non-gravitational fields are supposed to exist only on branes embedded in 5D. Accordingly, all the non-gravitational physical fields are represented as a $\delta$-function singularity in the energy-momentum tensor [6]. Similarly in the braneworld scenario [4][5], 10D supergravity can be embedded into 11D supergravity, again within certain $\delta$-function type singularities. The common feature between the braneworld scenarios and our result is that the $\delta$-function singularities represent important physical entities. Therefore it is natural to identify the singularities in the energy-momentum tensor in certain axisymmetric exact gravitational solutions in vacuum with $\delta$-function type closed string mass distributions.

In the cases of $\delta \geq 2$ of TS solutions, there are in total $\delta$ ring singularities, some of which may be completely hidden by event horizons, in addition to 'naked' ones lying outside of non-singular event horizons [10][11]. Our original motivation for studying the TS solutions with $\delta \geq 2$ was that these naked singularities are more 'physical' than the hidden ones, because they are more likely to be actual closed string-like mass distribution. However, in our actual computation (3.16) for the identification $P^0 = m$, we did not distinguish them. This is due to the philosophy that it is the total mass $m$ that should agree with the asymptotic mass $m$ [16], which of course includes those mass distributions *inside* event horizons. Or more intuitively, any mass distribution, even if it is hidden inside of event horizon, is 'visible' from outside the horizon. To put it differently, we know as a simple fact that even a point mass inside the event horizon in Schwarzschild metric definitely attracts a test particle outside of the event horizon. From these viewpoints, there is nothing contradictory in our prescription of identifying all the ring singularities. It is also true that the cases of $\delta \geq 2$ [10][11] have more 'naked' closed string-like mass distributions which are much more 'real' than those hidden by event horizons.

Despite the fact that we may need more supporting evidence based on alternative regu-larizations, our first results provide strong evidence of closed string like mass distribution for axially symmetric exact solutions. In other words, the general TS solutions can be regarded as classical backgrounds around closed strings. Since the distortion parameter $\delta$ corresponds also to the number of the ring singularities, *i.e.,* the number of rings, it is legitimate to regard $\delta$ as nothing other than the number of real closed strings with $\delta$-function type mass singularities. We believe that the encouraging results in this paper provide an entirely new directions to be explored both in the context of superstring theory and general relativity.

## References

[1] K. Schwarzschild, Sitzungsberichte Preuss. Akad. Wiss., **424** (1916).

[2] M. Green, J.H. Schwarz and E. Witten, *'Superstring Theory'*, Vols. **I** and **II**, Cambridge University Press (1987).

[3] J. Chazy, Bull. Soc. Math. France, **52** (1924) 17; H.E.J. Curzon, Proc. London Math. Soc. **23** (1924) 477.

[4] P. Hořava and E. Witten, Nucl. Phys. **B460** (1996) 506; Nucl. Phys. **B475** (1996) 94.

[5] N. Arkani-Hamed, S. Dimopoulos and G. Dvali, Phys. Lett. **429B** (1998) 263; I. Antoniadis, N. Arkani-Hamed, S. Dimopoulos and G. Dvali, Phys. Lett. **436B** (1998) 257.

[6] L. Randall and R. Sundrum, Phys. Rev. Lett. **83** (1999) 3370; Phys. Rev. Lett. **83** (1999) 4690.

[7] R.P. Kerr, Phys. Rev. Lett. **11** (1963) 237.

[8] A. Ya Burinskii, Phys. Lett. **A185** (1994) 441; *'Complex String as Source of Kerr Geometry'*, hep-th/9503094; Phys. Rev. **D57** (1998) 2392; *'Structure of Spinning Particle Suggested by Gravity, Supergravity & Low-Energy String Theory'*, hep-th/9910045, Czech. J. Phys. **50S1** (2000) 201.

[9] *See, e.g.,* A. Sen, Mod. Phys. Lett. **A10** (1995) 2081; P.H. Frampton and T.W. Kephart, Mod. Phys. Lett. **A10** (1995) 2571; A. Strominger and C. Vafa, Phys. Lett. **379B** (1996) 99; K. Behrndt, Nucl. Phys. **B455** (1995) 188; J.C. Breckenridge, D.A. Lowe, R.C. Myers, A.W. Peet, A. Strominger and C. Vafa, Phys. Lett. **B381** (1996) 423; C. Callan and J. Maldacena, Nucl. Phys. **B472** (1996) 591; G. Horowitz and A. Strominger, Phys. Rev. Lett. **77** (1996) 2368; J.M. Maldacena, *'Black Holes in String Theory'*, Ph.D. Thesis, hep-th/9607235; A. Dabholkar and J.A. Harvey, Phys. Rev. Lett. **63** (1989) 478; A. Dabholkar, G.W. Gibbons, J.A. Harvey and F. Ruiz Ruiz, Nucl. Phys. **B340** (1990) 33; C.G. Callan, Jr., J.M. Maldacena, A.W. Peet, Nucl. Phys. **B475** (1996) 645.

[10] A. Tomimatu and H. Sato, Prog. Theor. Phys. **50** (1973) 95.

[11] M. Yamazaki and S. Hori, Prog. Theor. Phys. **57** (1977) 696; Erratum-ibid. **60** (1978) 1248; S. Hori, Prog. Theor. Phys. **59** (1978) 1870, Erratum-ibid. **61** (1979) 365.

[12] H. Nishino, Phys. Lett. **359B** (1995) 77.

[13] H. Weyl, Ann. de Phys. **54** (1917) 117.

[14] J.M. Bardeen, Astrophys. Jour. **162** (1970) 71.

[15] D. Kramer, H. Stephani, E. Herlt and M. MacCallum, *'Exact Solutions of Einstein's Field Equations'*, Cambridge University Press (1980).

[16] R. Arnowitt, S. Deser and C. Misner, in *'Gravitation'*: *'An Introduction to Current Research'*, ed. L. Witten (New York, Wiley, 1962).